\documentclass[conference]{IEEEtran}
\IEEEoverridecommandlockouts
\usepackage{cite}
\usepackage{amsmath,amssymb,amsfonts}
\usepackage{algorithmic}
\usepackage{graphicx}
\usepackage{textcomp}
\usepackage{xcolor}

\usepackage{algorithm}
\usepackage{algorithmic}
\usepackage{amsmath}
\usepackage{url}
\usepackage{color}
\usepackage{multirow}
\usepackage{graphicx}
\usepackage{subfig}
\usepackage{epstopdf}
\DeclareGraphicsExtensions{.pdf,.jpeg,.png,.eps}
\usepackage{listings}
\usepackage{color}
\usepackage{setspace}

\def\BibTeX{{\rm B\kern-.05em{\sc i\kern-.025em b}\kern-.08em
    T\kern-.1667em\lower.7ex\hbox{E}\kern-.125emX}}
\begin{document}

\title{swCaffe: a Parallel Framework for Accelerating Deep Learning Applications on Sunway TaihuLight\\
\thanks{}
}

\author{\IEEEauthorblockN{Jiarui Fang*, Liandeng Li*}
\IEEEauthorblockA{{*Equal Contribution} \\
\IEEEauthorblockA{\textit{Tsinghua University}}
\IEEEauthorblockA{\textit{National Supercomputing Center in Wuxi }}
\{lld14, fjr14\}@mails.tsinghua.edu.cn}
\and
\IEEEauthorblockN{Haohuan Fu}
\IEEEauthorblockA{\textit{Tsinghua University} \\
\textit{National Supercomputing Center in Wuxi } \\
haohuan@mail.tsinghua.edu.cn
}
\and
\IEEEauthorblockN{Jinlei Jiang}
\IEEEauthorblockA{\textit{Tsinghua University} \\
jjlei@tsinghua.edu.cn
}
\and
\IEEEauthorblockN{Wenlai Zhao, Conghui He}
\IEEEauthorblockA{\textit{Tsinghua University} \\
\textit{National Supercomputing Center in Wuxi } \\
cryinlaugh@126.com, heconghui@gmail.com
}
\and
\IEEEauthorblockN{Xin You}
\IEEEauthorblockA{\textit{Beihang University}\\
youxin2015@buaa.edu.cn
}
\and
\IEEEauthorblockN{Guangwen Yang}
\IEEEauthorblockA{\textit{Tsinghua University},\\
\textit{National Supercomputing Center in Wuxi }\\
}
ygw@mail.tsinghua.edu.cn
}

\maketitle

\begin{abstract}
This paper reports our efforts on swCaffe, a highly efficient parallel framework for accelerating deep neural networks (DNNs) training on Sunway TaihuLight, the current fastest supercomputer in the world that adopts a unique many-core heterogeneous architecture, with 40,960 SW26010 processors connected through a customized communication network.
First, we point out some insightful principles to fully exploit the performance of the innovative many-core architecture.
Second, we propose a set of optimization strategies for redesigning a variety of neural network layers based on Caffe.
Third, we put forward a topology-aware parameter synchronization scheme to scale the synchronous Stochastic Gradient Descent (SGD) method to multiple processors efficiently.
We evaluate our framework by training a variety of widely used neural networks with the ImageNet dataset. 
On a single node, swCaffe can achieve 23\%\~{}119\%  overall performance compared with Caffe running on K40m GPU.
As compared with the Caffe on CPU, swCaffe runs 3.04\~{}7.84x faster on all the networks.
Finally,  we present the scalability of swCaffe for training of ResNet-50 and AlexNet on the scale of 1024 nodes.

\end{abstract}

%
%
%


\section{Introduction}
Deep Learning \cite{lecun2015deep} has already proven its usability in a variety of applications \cite{najafabadi2015deep}. In order to achieve better result or to deal with more complex problems, the scale of network gets larger and larger. 
As large network structures require substantial computational power, memory throughput and storage capacity, training neural networks poses a great challenge to the underlying hardware.
Since single processor efficiency has reached the physical limits of the hardware, scaling DNN training over parallel supercomputer becomes a good solution to satisfy the computation and storage requirements.

Sunway TaihuLight \cite{fu2016sunway}, a supercomputer that ranks first in the world currently, is powered by the SW26010 many-core processors with a total computing capacity of over 100 PFlops. The SW26010 processor is designed with on-chip heterogeneous techniques and provides a peak double-precision performance of 3.02 TFlops. Over 40,000 SW26010 processors are connected hierarchically with high-bandwidth customized hierarchical network.

Our previous work \cite{fang2017swdnn} has already explored the possibility of developing highly efficient convolution subroutines on SW26010. 
However, there remains great challenges to scale the entire DNN training to larger clusters.  
First, as main-stream DNN frameworks are usually designed for CPU-GPU hybrid system, straight-forward migrations or implementations of these frameworks to the brand new architecture can not achieve satisfactory performance. 
Redesigning a variety of basic DNN layers according to the characteristics of the SW26010 processors is the only way to unleash the potential performance of the supercomputer. 
Second, parallel training suffers from frequent communications and imbalanced operations among a large number of nodes. 
A customized communication strategy is necessary to take advantage of the network topology of Sunway TaihuLight.
Third, the parallel disk I/O of the input data can also become a bottle-neck in large-scale DNN training.

To solve the above challenges and facilitate network training tasks on TaihuLight, we redesign the widely-used Caffe framework and customize a set of routines to best fit the unique heterogenous architecture of SW26010, and further scale it to a large number of nodes. Our main contributions are as follows:
\begin{itemize}
\item We point out a set of general principles for designing parallel algorithm that fit the different aspects of SW26010 hardware characteristics.
\item A Caffe-based framework for SW26010 processor, namely swCaffe, is developed. It incorporates a set of optimization strategies and redesigns a variety of DNN layers to fully squeeze every bit of performance from the SW26010 processors.
\item We put forward a parallel synchronous SGD method to scale swCaffe on multiple nodes with highly-efficient parameter synchronization and parallel I/O strategy.
\item The swCaffe is open-sourced on \cite{swcaffe}, which maintaining the same interfaces as Caffe but can be deployed more efficiently on the TaihuLight system.
\end{itemize}

The rest of the paper is organized as follows.
In Section \ref{sec:background}, we describe Sunway TaihuLight architecture and DNN training methods as backgrounds.
In Section \ref{sec:1node}, we describe the principles for parallel algorithm design on SW26010 and optimization methods of swCaffe for DNN layers based on these principles.
In Section \ref{sec:scale}, we present our methodology to scale swCaffe on multiple nodes.
In Section \ref{sec:conlusion}, we conclude with a brief discussion of future work.

\section{Background}
\label{sec:background}

The Sunway TaihuLight supercomputer is composed of 40,960 nodes with a total of 10,649,600 cores. The nodes are connected through a customized network.

\subsection{SW26010 Many-core Processor}
\begin{figure}[ht!]
\centering
\includegraphics[width=0.45\textwidth]{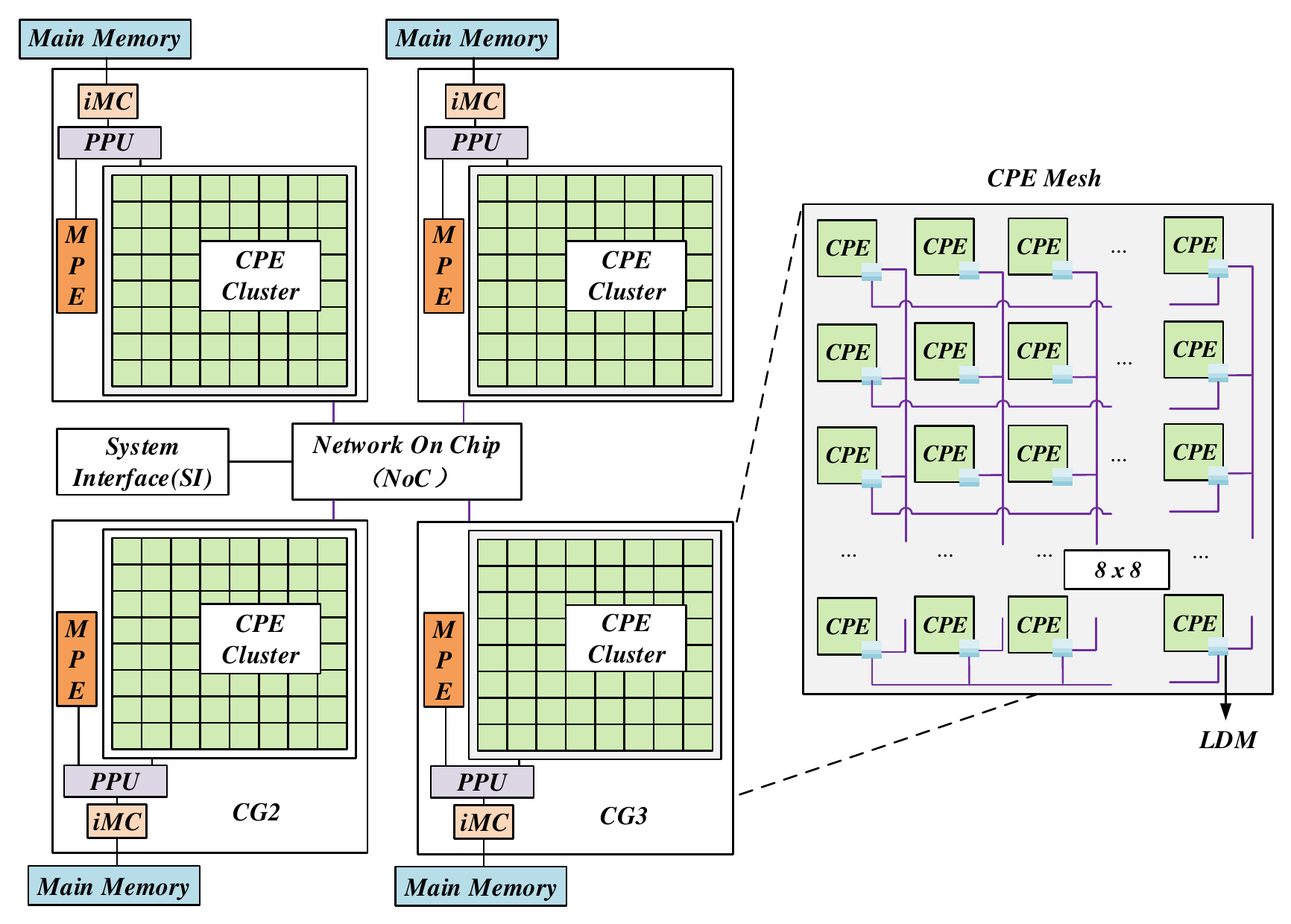}
\caption{The architecture of SW26010 many-core processor}
\label{fig:conv}
\end{figure}

The general architecture of the SW26010 is shown in Figure \ref{fig:conv}. The SW26010 processor includes 4 core-groups (CG) connected via the network on chip (NoC). Each CG includes one management processing element (MPE), one computing processing element (CPE) cluster with 8x8 CPEs, and one memory controller (MC). The processor connects to other outside devices through a system interface (SI).

Each group has its own memory space (8GB DDR3 memory for each), which is connected to the MPE and the CPE cluster through the MC. Four core groups connect to four 128-bit DDR3 memory controllers with a theoretical memory bandwidth of 136GB/s.

The MPE and CPE are both 64-bit RISC cores, which are running at 1.45 GHz with 256-bit SIMD instructions supported. Each MPE has a 32 KB L1 data cache, a 32 KB L1 instruction cache, and a 256 KB L2 cache while each CPE has a 16KB instruction cache and a 64 KB local directive memory (LDM), also known as Scratch Pad Memory (SPM), which should be explicitly controlled by user.

The 8$\times$8 CPEs are able to communicate with each other via register buses. CPEs that fall into the same row or same column can send messages each other through the fast register communication mechanism. In one cycle, the registers support up to 256-bit broadcast or P2P communication between two CPEs.

\subsection{Network Topology of Sunway TaihuLight}
\label{sec:network}
The customized network of TaihuLight is divided into 2 levels, namely a fat tree at the top and a supernode network at the bottom. 
The central switching network is responsible for communicating different supernodes, which is designed to use only a quarter of the potential bandwidth instead of a fully connected network. 
Each supernode has 256 nodes connected by high bandwidth network using the static destination-based strategy as its route policy.
TaihuLight uses FDR 56Gbps network interface cards (NICs) and provides a 70TB/s bisection network bandwidth in total. 
The theoretical bandwidth between any two nodes is 16GB/s. However, it only achieves 12GB/s with a latency at the level of micro-second when nodes are communicated with the Message Passing Interface (MPI).


\subsection{DNN Training Process and Frameworks}
Deep learning is used to solve the following optimization problem.
\begin{equation}
\mathop{\arg\min}\limits_{\theta} \  f(\theta) = \frac{1}{N}{\sum_{n=1}^{N}f_n(\theta)}
\end{equation}
where $\theta$ is the model parameters (or weights) we are looking for; $N$ is the number of samples; $f(\theta)$ is typically in a form of a DNN; $f_n(\theta)$ is the loss function of the $n^{th}$ sample. The stochastic gradient descent (SGD) method is the de facto method for DNN training.

A typical implementation of SGD is iterating the forward-backward propagations. 
The forward propagation step uses a mini-batch of training data as input to calculate the intermediate activations after each layers, while the backward propagation step uses the intermediate activations to perform gradient computation. The gradient to model parameters are then applied to the model after each backward propagation step.

Caffe \cite{jia2014caffe} is an open-sourced software framework used for DNN training. It is written in C++ and widely adopted in research experiments and industry deployments. 
Caffe implements DNN training with three major components, namely layers, net and solvers, corresponding to three optimization levels. 
Layers implement the algorithm of different neural network layers, related with the algorithm level optimization targeting different underlying hardware and platforms.
The net defines the network structure of a DNN model and implements the forward and backward propagations, so it allows optimizations for the process of one training iteration, such as process parallelization and memory optimizations.
Solvers control the network training process and implement the parameter tuning algorithms, such as Stochastic Gradient Descent (SGD). Therefore, optimizations for network training algorithms and distributed training process should be involved in the solvers.
The original Caffe framework is designed for standalone training with one HPC server, and only supports GPU accelerators. In order to efficiently map the framework onto Sunway TaihuLight supercomputer, we need to refactor or redesign the implementation of the above three components, so as to fit the unique architecture of the processors and to support distributed training over multiple nodes.

\section{Design and Implementation of DNN Framework on SW26010}
We first present principles of parallel algorithm design on SW26010 and then introduce our strategies to redesign the computing kernels of different DNN layers on SW26010 under the guidelines of these principles.

\subsection{Principles of Parallel Algorithm Design on SW26010}
The SW26010 is a brand new processor, which is totally different from other many-core processors used for DNN training, such as GPU and Intel Xeon Phi co-processors. Table \ref{table:manycores} shows the comparison of different aspects among SW26010, GPU and KNL. The methodologies for accelerating neural layers in main-streaming architectures (GPU, KNL) are not suitable for the SW26010 architecture. It often results in extremely poor performance if we migrate the framework that runs on GPU or KNL to SW26010 in a straight forward way.

A clear understanding of the advantages and disadvantages of the hardware architecture is of great importance to fully squeeze every bit of potential performance from Sunway TaihuLight. As a result, we propose the a set of principles as the guidelines when desiging the high performance applications.

\begin{table}
\small
\caption{Comparison of SW, Intel Knight Landing (KNL) and NVIDIA K40m processors}
\begin{tabular}{c||c|c|c}
\hline
Specifications  & SW26010 & Nvidia K40m & Intel KNL \\
\hline
\hline
Release Year & 2014 & 2013 & 2016\\
\hline
Bandwidth(GB/s) & 128 & 288 & 475 \\
\hline
float perf. (TFlops ) & 3.02 & 4.29 & 6.92\\
\hline
double perf. (TFlops ) & 3.02 & 1.43 & 3.46\\
\hline
\end{tabular}
\label{table:manycores}
\end{table}

\textbf{Principle 1: Fully utilize $8\times8$ CPE mesh for computation-intensive tasks.}
The CPE cluster provides the computing capacity of 742.4 GFlops while the MPE only 11.6 GFlops in each CG theoretically. So the most important step to improve the performance is to offload the computationally intensive kernels to the $8\times8$ CPE mesh. Different levels of parallelism can also be carefully exploited within CPE clusters:

\begin{itemize}
  \item The parallelism between 64 CPEs is exploited by orchestrating data-independent tasks on each CPE simultaneously.
  \item For each CPE, data-level parallelism can be exploited by using 256-bit vector registers for SIMD operations.
  \item In addition, we can exploit instruction-level parallelism from two instruction pipelines, the floating-point pipeline and the memory access pipeline. Both pipeline issues instructions in order on their own pipeline, while independent instructions on different pipelines could be issued out of order.
\end{itemize}

\begin{figure}[ht!]
\centering
\includegraphics[width=0.5\textwidth]{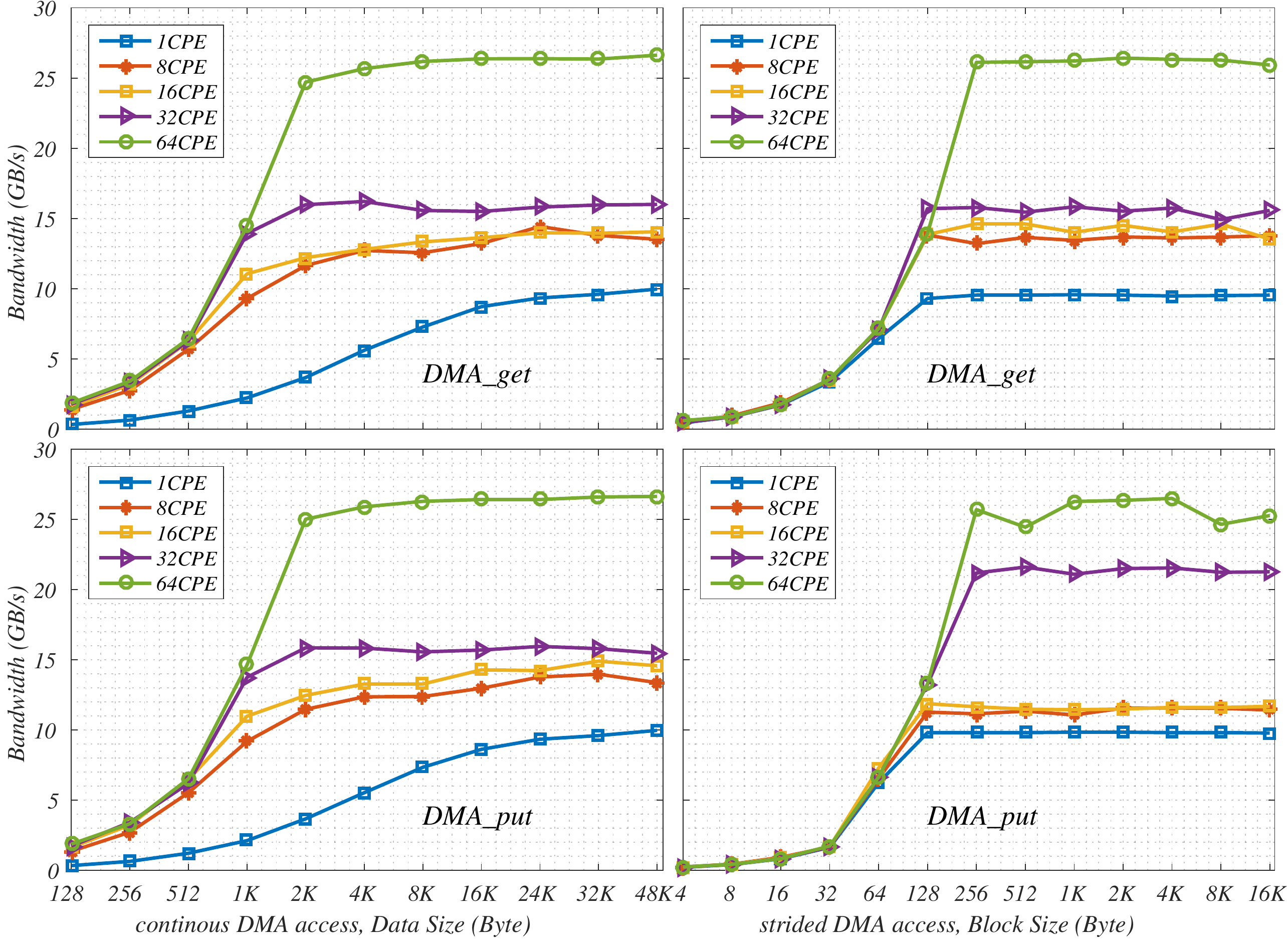}
\caption{Bandwidth of DMA get and put operations for continuous and strided data access patterns. 
The left two figures show how bandwidth varies with the data sizes of continuous DMA access for each CPE.
The right two figures show how bandwidth varies with the data block sizes of strided DMA access for each CPE, when the total accessed data size of each CPE is 32 KB.}
\label{fig:ldmbandwidth}
\end{figure}

\textbf{Principle 2: Always use LDM as intermediary cache for data movements between DDR3 memory.}
In each CG, the memory controller is responsible for connecting both the MPE and the CPE cluster to the DDR3 memory, which means the MPE and the CPEs share the theoretical memory bandwidth of 32 GB/s.
According to the benchmark in Figure \ref{fig:ldmbandwidth}, the DMA bandwidth saturates around 28 GB/s for both read and write.
However, the memory bandwidth between Memory-to-MPE and Memory-to-LDM is extremely different. The bandwidth of copying data from one DDR3 memory space to another through Memory-to-MPE is only 9.9 GB/s.
As a result, it is always preferred to use LDM as the intermediary cache, other than  accessing main memory from CPEs directly.

\textbf{Principle 3: Increase available memory bandwidth by transferring large data blocks.} 
The limited aggregated memory bandwidth and the high-computing power lead to an extremely high flop-per-byte ratio, which is $\frac{742.4 Gflops}{28 GBps} = 26.5$, compared with ratios of 14.90 and 14.56 for K40m and KNL, respectively.
To achieve satisfactory DMA bandwidth, we should keep in following points in mind during algorithm design.
First, data transfer should be conducted with 64 CPEs together.
Second, memory access from the CPE cluster in small granularity should be avoided as much as possible.
Size of data to be transferred for each CPE should be larger than 2 KB so that data transfer time can hide the hundreds of cycles LDM transfer latency.
Data block size for strided access should be at least 256 bytes so as to achieve satisfactory bandwidth performance.

\textbf{Principle 4: Reduce memory access by register-level communication among CPEs.}
Besides increasing available bandwidth, we can also reduce the amount of data transfer between LDM and memory to improve performance.
The register-level communication (RLC), which enables P2P/broadcast 256-bit data communications at the register level among CPEs, is a unique hardware characteristic of SW26010.
Direct RLCs are allowed only between CPEs within the same row or the same column, following an anonymous producer-consumer pattern with FIFO sending/receiving buffers (i.e., the send instruction is asynchronous, and the sender/receiver gets stalled if the sending/receiving buffer is full/empty).
If RLC transfers are fully pipelined, the overall P2P and broadcast bandwidth can reach 2549 GB/s and 4461 GB/s respectively \cite{xu2017benchmarking}. In this way, we can reuse the data in other LDMs on the same row/column in the CPE cluster to reduce bandwidth requirements between the main memory and LDMs.

\section{Parallel design of DNN layers}
\label{sec:1node}
A Deep Neural Network consists of different layers.
We present our optimization methods for the most frequently used layers in DNN applications, according to the principles pointed out in the previous section.

\subsection{Matrix-Multiplication Layer}
\label{sec:gemm}
The inner-product layers and other more complicated layers, such as Long Short Time Memory (LSTM) layers, are mainly involving General Matrix to Matrix Multiplication (GEMM) operations.
If data locality is fully exploited and near optimal memory bandwidth is achieved, GEMM operations can be implemented with a high flop-to-byte ratio.
To implement it on CPE cluster, we use the register communication proposed in \cite{fang2017swdnn}\cite{jiang2017towards} to increase data locality in LDM.
Assume we intend to perform GEMM operation $C += A \times B$, where matrix $A$, $B$ and $C$ are of sizes $m \times k$, $k \times n$, $m \times n$, respectively and can all fit into the 64 KB LDM.
Matrices are evenly divided to dimension of size $m/8, n/8$ and $k/8$.
A CPE is responsible for computing $m/8 \times n/8$ block of $C$ requiring an $m/8 \times k$ tile of $A$ and a $k \times n/8$ tile of $B$.
Note that, in this case, $7/8$ of both tiles of $B$ and $C$ required by this CPE are resident on remote LDM of other CPEs.
According to \textbf{Principle 4}, we can take advantage of the row and column register communication scheme to fetch remote data, as CPEs in the same row of the cluster share the tile of $A$, and CPEs in the same row of the cluster share the tile of $B$.

\begin{figure}[ht!]
\centering
\includegraphics[width=0.40\textwidth]{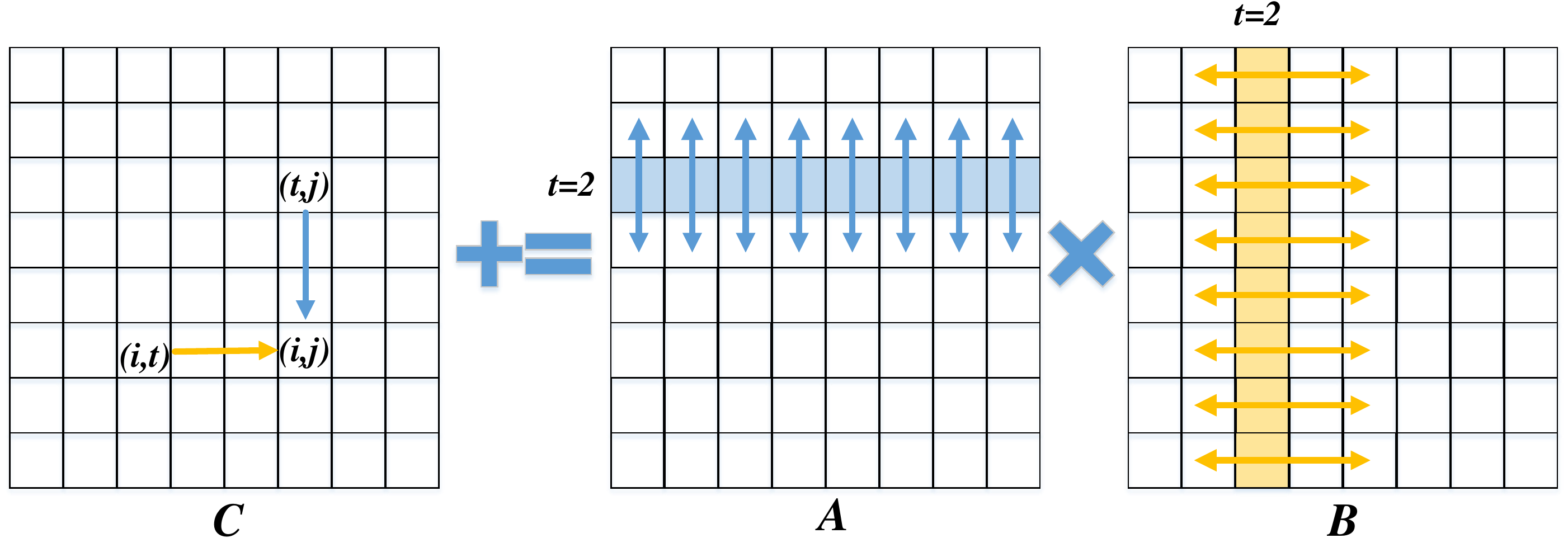}
\caption{Illustration of GEMM with register communication on CPE cluster.}
\label{fig:gemm}
\end{figure}

The GEMM operation can be finished in 8 steps as $C(i, j) += \sum_{t=0}^{7}A(i, t) \times B(i, t)$. (i, j) indicates the coordinate of the CPE, where data is resident, in the $8 \times 8$ cluster.
For each time step $t ( 0 \leq t  \leq 7)$, CPE$(i, t)$ loads data of $A(i, t)$ from LDM and broadcasts the data to other CPEs in the same column by column register communication.
Similarly, CPE$(t, j)$ loads data of $B(t, j)$ from LDM and broadcasts the data to CPEs in the same row.
Thus, CPE$(i,j)$ can receive both data of CPE$(i, t)$ and CPE$(t, j)$ and the computation of $C (i, j)+ = A (i, t) \times B (t, j)$ can be done in each time step.
Figure \ref{fig:gemm} illustrates the register communication operations when $t$ is 2.
This is optimal design with highest flop-to-byte ratio, as we only require to fetch matrices from memory to LDM once.

Blocking techniques are applied to matrices which are too large to fit into the LDM.
As the memory-LDM bandwidth is critical for the GEMM performance, 
the continuous data sizes of matrix blocks each CPE accesses should be large enough according to \textbf{Principle 3}.
As a result the dimension size of matrices should be large enough for good memory bandwidth.

SW26010 provides no inherent support for single-precision floating point operations, which is default precision option used in DNN.
As there is no instruction to support RLC for single precision data in the instruction set of SW26010, we always perform RLC operations with double-precision data and we conduct inline transformation for elements between double-precision to single-precision with SIMD instructions.

\subsection{Convolutional Layer}
\label{sec:conv}
The convolutional layers are the most compute-intensive parts when training Convolutional Neural Networks (CNNs).
Both time-domain methods with GEMM operations \cite{chetlur2014cudnn} and frequency-domain methods with FTT operations \cite{collobert2011torch7} are proposed to optimize convolutional layers on GPU.
Because GEMM operations can be perfectly optimized on CPE cluster with the register-level communication as mentioned previously, we adopt time-domain transformation methods.
To support different convolutional layer parameter configurations in real CNN applications, we propose a mixed strategy combining the explicit GEMM plan used in original Caffe and the implicit GEMM plan proposed in \cite{fang2017swdnn}.

\subsubsection{Explicit GEMM transformation}
To map convolution operations to GEMM and reuse the GEMM routine mentioned in Sec.\ref{sec:gemm}, we adopt an explicit GEMM transformation proposed for original Caffe.
In this case, input tensors are first transferred into matrices by im2col (image-to-column) operations before leveraging GEMM operations during forward propagation, while col2im (column-to-image) operations are performed after GEMM operations during backward propagation.
Assuming a convolutional layer has filter of size $(N_o, N_i, K, K)$, im2col operation transfers a 3D multi-channel image tensor of size $(C_i, R_i, N_i)$ to a 2D matrix of size $(C_o\times R_o, K\times K\times N_i)$.
$C_{i/o}$ and $R_{i/o}$ are column and row of output image, where $C_o = (C_i-K)/S+1$, $R_o = (R_i-K)/S+1$, where $S$ is the convolution stride.
$N_i$ is input channel number.
$N_o$ is filter channel number.
$K$ is filter size.
The dimension of batch-size $B$ is also introduced for blocking, which brings more optimization space for GEMM blocking.

\begin{figure*}[htb]
\centering
\includegraphics[width=0.85\textwidth]{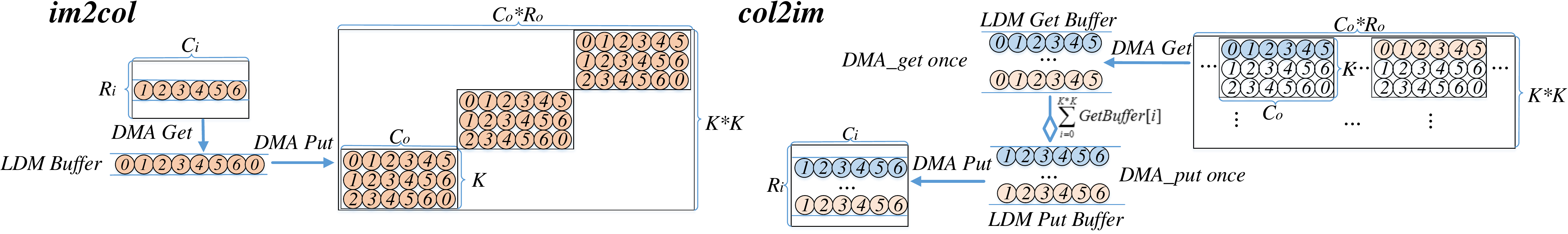}
\caption{Optimization for memory access for Im2col and col2im operations on one CPE.}
\label{fig:im2col}
\end{figure*}

As the filter tensor can be viewed as a matrix of size $(N_o, K\times K\times N_i)$, GEMM operation is performed on two matrices with common dimension of size $K\times K\times N_i$.
Im2col and col2im consist of irregular memory access pattern.
The convolutional layers in backward propagation can transfer matrix back to tensor with col2im, which has a reverse memory movement.
As indicated by \textbf{Principle 4}, irregular memory access of im2col and col2im should be implemented with DMA on CPE cluster.
Figure  \ref{fig:im2col} shows our im2col and col2im plan on one CPE.
During im2col process, each CPE reads one row of a input image into LDM buffer with DMA get operation.
After adding with pad, each CPE writes $K\times K$ line of data into memory.
Block sizes are critical for memory bandwidth in GEMM operation.

\subsubsection{implicit GEMM transformation}
The time overheads of im2col and col2im are not negligible for some layers.
An implicit GEMM transformation proposed in our previous work \cite{fang2017swdnn} is integrated to implement convolutional layers for swCaffe by blocking on dimensions of image width and input and output channels to increase data reuse in LDM.
However, when the input and output filter channel numbers are smaller than 64, performance of implicit method would largely degrade, because the amount of data in LDM with small channels is not large enough to support 256-bit SIMD and register communication operations.


Real applications apply convolutional layers with input images after zero padding.
Considering padding operation has already been implemented combining with im2col/col2im operations in explicit scheme,
we also propose a padding optimization in implicit GEMM transformation convolution layers by use a coordinate mapping techniques to avoid explicitly padding operations.
Details of padding and more other optimization techniques for convolutional layers can be found in our technique report released with source code \cite{swcaffe}.

\subsection{Tensor Transformation Layer}
The data of explicit GEMM transformation and implicit GEMM transformation are arranged differently.
In the explicit GEMM transformation plan, input and output tensors are of shape $(B, N, R, C)$ and filters are of shape $(N_o, N_i, K, K)$, which is also the default data layout for other layers.
In the implicit GEMM transformation plan, input and output tensors are of shape $(R, C, N, B)$ and filters are of shape $(K, K, N_o, N_i)$.
Note that the convolutional layers that can be accelerated with implicit transformation are gathered together.
The filters are local variables of this layers and its layout do not effect other layers.
In swCaffe, we add a tensor transformation layer, which has an 4D tensor input and an 4D tensor output with dimensions transposition between two different data layouts.

The tensor transformation in trans\_layer is mainly irregular memory movement and should also be accelerated on CPE cluster.
Stride DMA access is adopted to access a block of tensor into LDM.
SIMD shuffle instructions are used to transform data after load data from LDM to registers.

\subsection{Pooling Layer}
The pooling layer partitions the input image into a set of non-overlapping tiles and, for each such sub-region, outputs the maximum or average values of elements inside.
Since pooling layers are featured with massive memory copy operations, they should be implemented with DMA operations on CPE cluster.
We design different movement strategies according to the sizes of input images.
Assuming the tile size is $K\times K$.
According to \textbf{Principle 3}, we should increase the continuous data size as much as possible for data blocks.
Most of times, each CPE is in charge of pooling operation for multiple $K$ rows of input image.
When $K$ rows of image can not be fitted in LDM, we load number of columns into LDM as large as possible.
In this case, the data needed by LDM is not continuously stored in memory and strided DMA is used to access it.

%

\section{Scaling DNN framework on the TaihuLight}
\label{sec:scale}
In this section, we describe our design to scale swCaffe on multiple processors.

\subsection{Optimization for Communication of Model Parameters}
In our work, we adopt a data parallel scheme with synchronous Stochastic Gradient Descend (SSGD) algorithm to scale swCaffe, which is widely adopted in HPC clusters and supercomputer systems \cite{goyal2017accurate}\cite{you2017scaling} considering the high quality of network and balanced performance per node.
There are mainly two methods to implement model parameter synchronization in SSGD.
One method is using the parameter servers \cite{dean2012large} as the intermediary which stores the parameters among several servers.
The parameter server scheme is unable to sufficiently exploit the bandwidth potential on fully-connected network infrastructure of Sunway Taihlight, 
since the processor has only one network port, thus, receiving gradients simultaneously from a large number of workers could potentially become a bottleneck in the parameter server design and bandwidth between workers are not fully used.
The other method is to perform all-reduce operations on the gradients among all nodes and to update the parameters on each node independently \cite{you2017scaling}.
We adopt the latter approach to take advantage of the MPI routines optimizing for the supercomputer system, as the former approach is designed for synchronization based on low-bandwidth network infrastructures, like Ethernet.
Our parallel synchronous SGD algorithm is described in Algorithm \ref{algo:distsgd}.
\begin{algorithm}
    \caption{Parallel SSGD algorithm on processor $k$}
    \label{algo:distsgd}
    \begin{algorithmic}[1]
     \renewcommand{\algorithmicrequire}{\textbf{Input:}}
     \renewcommand{\algorithmicensure}{\textbf{Output:}}
     	\REQUIRE dataset $\chi$, mini batch size $b$ per CPU, the number of node $N$, initial learnable parameters $w = {w[0], ... , w[M]}$
	    \STATE launch 4 threads on 4 CGs
            \FOR {$t = 0,1, ... max\_iter$}
		\FOR {each threads $i$}
            		\STATE sample $\frac{1}{4}$ mini-batch ($\frac{b}{4}$ elements) as $x_i$ from $\chi$
			\STATE calculate $\nabla f(x_i; w_t)$ with forward and backward propagation
		\ENDFOR
		\STATE threads\_synchronization()
		\STATE $G_t^k = \frac{1}{4}\sum_{i=1}^{4}\nabla f(x_i; w_t)$
		\STATE $All$-$reduce$ $G_t^k : G_t \gets \frac{1}{N}\sum_{k=1}^{N}{G_t^k}$
		\STATE $w_{t+1} \gets SGD(w_t, G_t)$
            \ENDFOR
            \STATE finalize 4 threads
    \end{algorithmic}
\end{algorithm}

\begin{figure}[ht!]
\centering
\includegraphics[width=0.5\textwidth]{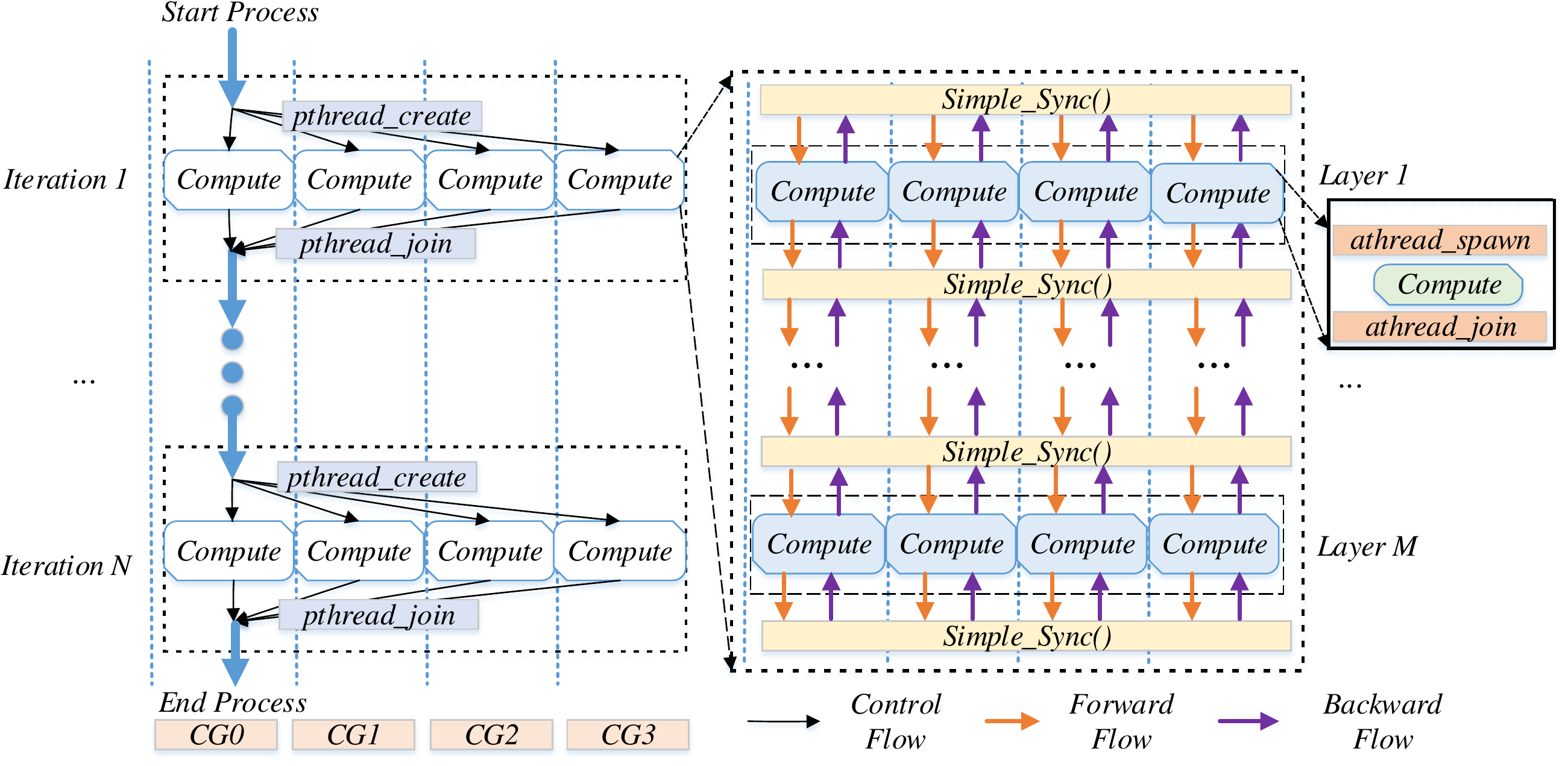}
\caption{Scaling DNN training with multi-threading technique on 4CGs inside one processor.}
\label{fig:4cg}
\end{figure}

As shown in Fig.\ref{fig:4cg}, we use multiple-threading technique among 4 CGs inside one processor to calculate the averages of gradients.
At the beginning of each iteration, we call pthread\_create() to start 4 threads on 4 CGs.
Each process is able to launch light-weight CPE threads to load work tasks onto CPE cluster, in order to perform forward-backward propagations of 1/4 of data in that mini-batch.
Afterwards, each CG achieves its local parameter gradients and CG 0 sums them together to achieve the average gradients of this mini-batch.
To synchronize the sub-threads, we implement a synchronization function by ourself, which is based on a handshake (initiation-confirmation) strategy through the semaphore stored in the shared memory.

\begin{figure}[ht!]
\centering
\includegraphics[width=0.5\textwidth]{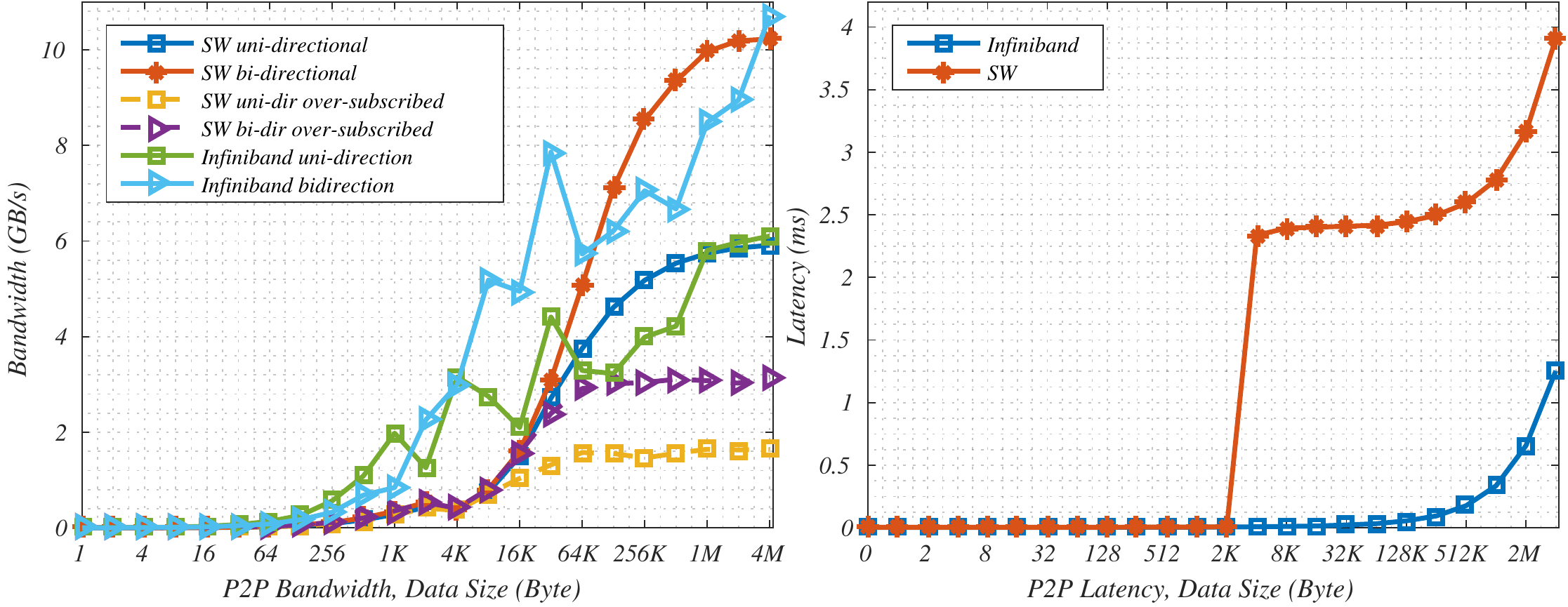}
\caption{Comparison of MPI P2P communication bandwidth and latency between the Sunway network and an Infiniband FDR network.}
\label{fig:mpibenchmark}
\end{figure}

To synchronize the gradients across nodes, we implement a customized all-reduce communication.
The default \texttt{MPI\_Allreduce} routine provided by compiler, which is modified from Open MPI\footnote{https://www.open-mpi.org/}, can not be directly applied for swCaffe for mainly three reasons.
First, the Sunway network is characterized by high latency, thus \texttt{MPI\_Allreduce} routines designed for low latency network hardware are no longer suitable in this situation.
As shown in Fig. \ref{fig:mpibenchmark}, we compare the Sunway network with an Infiniband FDR network.
While achieving similar high-bandwidth as Infiniband, the Sunway network has higher latency when message size is larger than 2KB.
Second, the communication pattern in \texttt{MPI\_Allreduce} is not aware of the topology of hierarchical network as mentioned in Sec. \ref{sec:network}.
If every node in one supernode performs point-to-point communication with a different node in another supernode, it will result in over-subscribed interconnect across supernodes.
As shown in Fig. \ref{fig:mpibenchmark}, the over-subscribed bandwidth between two supernodes is around $\frac{1}{4}$ of full bandwidth.
Third, the sum operation after data gathering in \texttt{MPI\_Allreduce} is performed on MPEs, thus it is not efficient in the case of large parameter amount.

We improve the all-reduce operation considering its high latency and topological property.
Before the introduction of our customized algorithms, we use the cost model proposed in \cite{thakur2005optimization} to evaluate our all-reduce in terms of latency and bandwidth use.
We assume that the time taken to send a message between any two nodes can be modeled as $\alpha + \beta {n}$, where $\alpha$ is the latency (or startup time) per message, independent of message size.
$\beta$ is the transfer time per byte, and $n$ is the number of bytes transferred.
More specifically, $\beta_1$ is the transfer time inside one supernode and $\beta_2 (\approx \frac{1}{4}\beta_1)$ is time across supernodes when bandwidth is over-subscribed.
In the case of reduction operations, we define $\gamma$ to be the computation cost per byte for performing the reduction operation locally on any node.
We also define $p$ to be the total number nodes in all-reduce operation and $q$ to be the number of nodes in one supernode.

\begin{figure*}[ht!]
\centering
\includegraphics[width=0.85\textwidth]{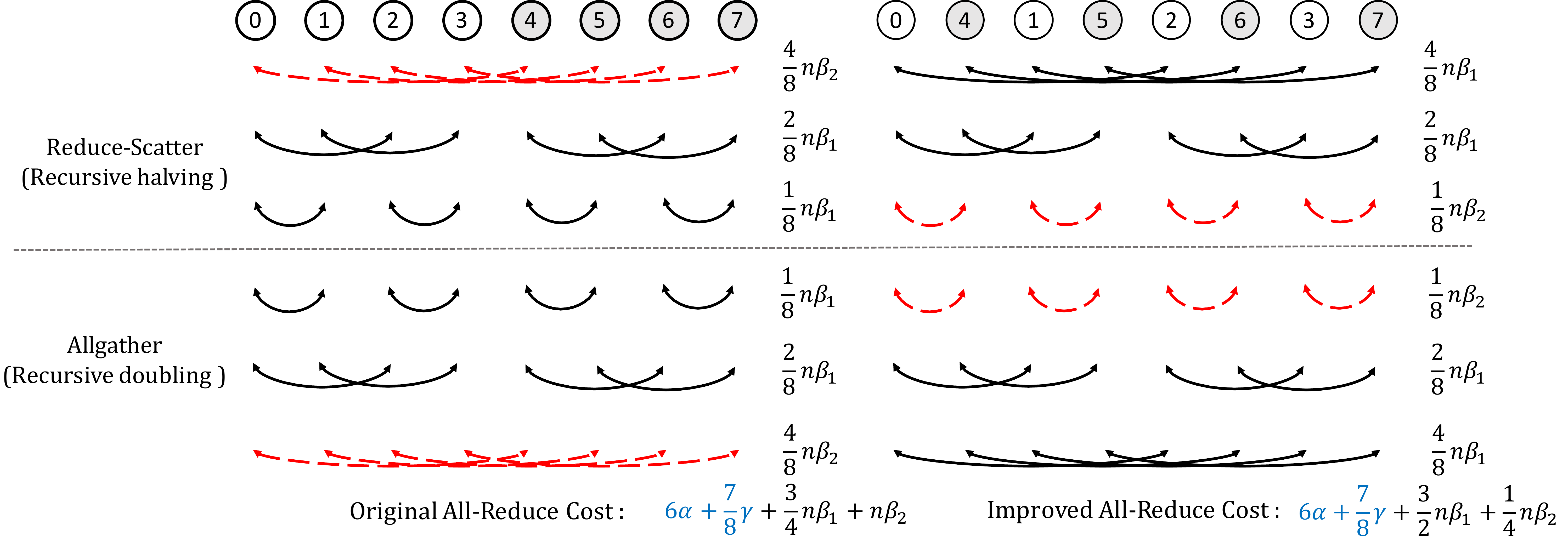}
\caption{A simple example with 8 nodes distributed in 2 supernodes to illustrate improvement of our all-reduce algorithm. Node 0-3 are in one supernode and node 4-7 are in the other one. Communication costs are labeled on right sides and dashed red lines indicate over-subscribed cross supernode communication. Our method can largely reduce the traffic across supernodes.}
\label{fig:allreduce}
\end{figure*}

Considering the high latency characteristics of the Sunway network, the popular ring-based algorithms \cite{patarasuk2009bandwidth}, having a $p\alpha$ latency term, are not our best candidates.
We choose a binomial-tree-based algorithm used in MPICH \cite{thakur2005optimization}, which has a $2\log{p}\alpha$ latency term, as our baseline to improve.
An all-reduce operation is implemented with an allgather phase after a reduce-scatter phase.
Instead of storing all results at the root node, reduce-scatter phase adopts the Recursive Halving algorithm to scatter reduction results among all nodes.
In the first step, each node exchanges $n/2$ data with a node that is a distance $p/2$ away.
Each node sends the data needed by all nodes in the other half, receives the data needed by all nodes in its own half,  and performs the reduction operation on the received data. 
In the second step, each node exchanges $n/4$ data with a node that is a distance  $p/4$ away.
This procedure continues recursively, halving the data communicated at each step, for a total of $\log p$ steps.
Recursive Doubling algorithm, analogous to the Recursive Halving algorithm, is adopted to collect partial results from other nodes for each node in the allgather phase.
In the first step, nodes that are a distance 1 apart exchange their  $n/p$ data.
In the second step, nodes that are a distance 2 apart exchange their own data as well as the data they received in the previous step, which has a size of $2{n}/{p}$ in total.
In the third step, nodes that are a distance 4 apart exchange their own data as well the data they received in the previous two steps, which has a size of $4{n}/{p}$ in total.
Nodes exchange message size up to $(2^{\log{p}-1}n)/p$ with the nodes that are a distance $p/2$ apart in the last step.
A simple example of such all-reduce implementation is illustrated on the left side of Fig. \ref{fig:allreduce}.

In the original implementation, nodes within the same supernode are assigned adjacent logical node numbers.
In the first several steps of Recursive Halving and last several steps of Recursive Doubling, each node has to communicate with a node far away in another supernode,
resulting in over-subscription between supernodes and achieves merely 1/4 of full bi-direction bandwidth.
The costs of original all-reduce are illustrated in Equ. \ref{equ:allreduce}, Equ. \ref{equ:rechav}, and Equ. \ref{equ:allgather}.
The last two equations are obtained by summing the costs for each time step, which can be viewed as a geometric progression. 
If $p$ is much larger than $q$, term $(p-q)\beta_2\frac{n}{p}$ will account for most of the communication time.

\begin{equation}
\label{equ:allreduce}
t_{allreduce} = t_{reduce-scatter} + t_{allgather}
\end{equation}
\begin{equation}
\label{equ:rechav}
t_{reduce-scatter} = \log{p}\alpha +  (q-1)\beta_1\frac{n}{p} + (p-q)\beta_2\frac{n}{p} + \frac{p-1}{p}n\gamma
\end{equation}
\begin{equation}
\label{equ:allgather}
t_{allgather} = \log{p}\alpha +  (q-1)\beta_1\frac{n}{p} + (p-q)\beta_2\frac{n}{p} 
\end{equation}

We notice that the communication traffic in different steps is not balanced.
Recursive Halving gradually reduces traffic, while Recursive Double gradually increases traffic.
Considering the topology of the Sunway network, a better all-reduce implementation should place heavy communication traffic inside one supernode and light one across supernodes.
We redesign the relationship between physical distance and logical distance used in all-reduce algorithm, by incrementally assigning logical numbers to nodes of different supernodes in a round robin way.
For example, assuming we have 4 supernodes,
Nodes numbered 0,4,8,... belong to supernode 0, nodes numbered 1,5,9,... belong to supernode 1, and so on.
As shown in Fig. \ref{fig:allreduce}, the new all-reduce conducts cross-supernode communication in the last $\log \frac{p}{q}$ steps of reduce-scatter phase and the first $\log \frac{p}{q}$ steps of allgather phase. For these steps, we only need to exchange relative small amount of message.
The new costs are shown in Equ. \ref{equ:newrechav} and Equ. \ref{equ:newallgather}.
As we can see, new implemenation largely reduces the coefficient of $\beta_2$ from $p-q$ to $\frac{p}{q}-1$, thus reducing the overhead caused by over-subscribed communication.

\begin{equation}
\label{equ:newrechav}
t_{new-reduce-scatter} = \log{p}\alpha +  (p-\frac{p}{q})\beta_1\frac{n}{p} + (\frac{p}{q}-1)\beta_2\frac{n}{p} + \frac{p-1}{p}n\gamma
\end{equation}
\begin{equation}
\label{equ:newallgather}
t_{new-allgather} = \log{p}\alpha +  (p-\frac{p}{q})\beta_1\frac{n}{p} + (\frac{p}{q}-1)\beta_2\frac{n}{p} 
\end{equation}

In addition, sum operations after data gathering are implemented on four CPE clusters of the processor.
The parameters of different layers can vary greatly in size.
In VGG-16, the first fully-connected layer is 102 MB, while the first convolutional layer is only 1.7 KB.
Sum operation for layer gradients of small parameter size can be inefficient, because we can not fully utilize the memory bandwidth to access data in small granularity.
We pack the gradients of all layers together to performance all-reduce after backward propagation.
Such scheme can fully utilize both network bandwidth for communication and memory bandwidth for sum operation.

\subsection{Parallel I/O optimization}
Computing nodes in Sunway TaihuLight adopt a shared file system.
Each worker of the parallel DNN training task uses an I/O thread to prefetch one mini-batch data via random sampling prior to each iteration.
The file system on Sunway TaihuLight adopts a single-split mode for data distribution by default, which indicates that one file will only be distributed on one disk array.
In this case, if we read the file concurrently, as the number of processes increases, the aggregate read bandwidth of multiple concurrent processes can quickly reach the upper limit of a single disk array.
As a result, each process will get a bandwidth drop and the entire reading time becomes longer.

We improve the aggregated bandwidth of disk arrays by increasing the number of stripe to 32 and modifying the splitting size to 256 MB.
Data is distributed on 32 disk array under the round robin strategy with block size as 256 MB.
Assume that one process is required to read a mini-batch data size of 256 for ImageNet images.
The data size for this mini-batch is around 192 MB.
Since each process always accesses consecutive 192 MB of data, a single process can access at most two disk arrays.
Accordingly, the number of processes required per disk array is also reduced to at most $N/ 32 \times 2$, where $N$ is the number of processes.

\section{Results}
We implement swCaffe with customized Sunway REACH (Open64 based) C compiler and SWMPI 2.2 (Mvapich 2.2 based) C++/MPI compileron TaihuLight.
We compare its performance with the original Caffe built with g++-4.8.0, CUDA-8.0 Tooltik and cuDNN-v5.1, and deployed on a hybrid system with an intel 12-core E52680 V3 CPU \footnote{E52680 V3's memory bandwidth is 68 GB/s and peak performance is 1.28 TFlops} equipped with a NVIDIA K40m GPU card. 
We conduct our experiments based on the public 1000-way ImageNet dataset \footnote{http://www.image-net.org/}.


\begin{table*}[ht!]
\scriptsize
\caption{Combination of explicit and Implicit GEMM transformation on one CG for Convolutional Layer in VGG-16 with batch-size = 128}
\begin{center}
\begin{tabular}{|c|c|c|c|c|c|c|c|c|c|c|c|c|c|}
\hline
\multirow{2}{*}{\textbf{conv}}  & \multirow{2}{*}{\textbf{Ni}} & \multirow{2}{*}{\textbf{No}} & \multirow{2}{*}{\textbf{Ci/Ri}} & \multicolumn{3}{|c|}{forward time(s)} &  \multicolumn{3}{|c}{weight\_diff backward(s)} & \multicolumn{3}{|c|}{in\_diff backward(s)} \\
 \cline{5-13}
 & 				  & 				      & 				 				      &  \textbf{implicit} & \textbf{explicit} & \textbf{Gflops} &  \textbf{implicit} & \textbf{explicit}  & \textbf{Gflops}  &  \textbf{implicit} & \textbf{explicit} & \textbf{Gflops} \\
 \hline
1\_1 	&	3	&	64	&	224	& $-$  &	\textbf{4.19}		&5.29 		& $-$  	&	\textbf{1.10}	&20.18 &	NA & NA & NA \\
 \hline
1\_2 		&	64	&	64	&	224	&	\textbf{4.30}		&	7.79  	&110.83 	&	$-$  	&	\textbf{5.22} &90.49 	& $-$ & \textbf{14.97} & 31.63\\
 \hline
2\_1 	&	64	&	128	&	112	&	\textbf{1.63}			 & {2.45}  		& 146.68 &	$-$  			&	\textbf{1.33} & 176.70 	& $-$ &	\textbf{3.61} & 65.65\\
  \hline
2\_2 	&	128	&	128	&	112	&	\textbf{2.34}			&	3.14		 & 202.52  &	2.26			&	\textbf{2.25}	& 209.26  &	\textbf{2.39}	& 6.11 & 198.41\\
   \hline
3\_1 	&	128	&	256	&	56	&	1.06	&	\textbf{0.73}	& 323.10 &	0.92	&	\textbf{0.68}	& 351.07 &	\textbf{0.95}	& 1.69 &248.92\\
    \hline
3\_2 	&	256	&	256	&	56	&	1.79	&	\textbf{1.14}	& 414.62 &	1.56	&	\textbf{1.29}	& 369.23 &	\textbf{1.82}	& 3.05 & 260.47\\
     \hline
3\_3 	&	256	&	256	&	56	&	1.79	&	\textbf{1.14}	& 415.97 &	1.56	&	\textbf{1.27}	& 376.02 &	\textbf{1.82}	& 3.03 & 260.46\\
      \hline
4\_1 	&	256	&	512	&	28	&	0.84	&	\textbf{0.69}	& 344.42 &	\textbf{0.70}	&	0.71	& 336.32 &	\textbf{0.85}	& 0.95 & 277.64\\
       \hline
4\_2 	&	512	&	512	&	28	&	1.68	&	\textbf{1.33}	& 347.36 &	\textbf{1.27}	&	1.33	& 372.75 &	\textbf{1.75}	& 1.89 & 270.54\\
        \hline
4\_3 	&	512	&	512	&	28	&	1.68	&	\textbf{1.33}	& 348.50 &	\textbf{1.27}	&	1.67	& 372.75 &	\textbf{1.75}	& 1.87 & 270.52\\
         \hline
5\_1 	&	512	&	512	&	14	&	\textbf{0.40}	&	0.62	& 293.58 &	\textbf{0.31}	&	0.65	& 376.94 &	\textbf{0.43}	& 0.80 & 274.26\\
          \hline
5\_2 &	512	&	512	&	14	&	\textbf{0.40}	&	0.63	& 293.58 &	\textbf{0.31}	&	0.78	& 376.94 &	\textbf{0.43}	& 0.84 & 274.26\\
           \hline
5\_3 	&	512	&	512	&	14	&	\textbf{0.40}	&	0.63	& 293.59 &	\textbf{0.31}	&	0.65	& 377.03 &	\textbf{0.43}	& 0.84 & 274.27\\
            \hline
\end{tabular}
\end{center}
\label{tab:conv}
\end{table*}

\subsection{Results for optimizations on different layers}
We analyze the performance of convolutional layers with both explicit and implicit GEMM transformation strategies proposed in Sec. \ref{sec:conv}.
Table \ref{tab:conv} presents the measured time and throughput for each convolutional layer of the VGG-16 \cite{simonyan2014very} network with batch-size 128.
VGG-16 has 12 convolutional layers and covers most commonly used parameter configurations.
In terms of the forwardprop in conv1\_1 and backwardprop in conv1\_1,conv1\_2 and conv2\_1, implicit strategy is unable to handle small channel sizes and explicit strategy is the only solution.
For most parameter configurations, implicit strategy outperforms explicit strategy.
However, explicit strategy is slightly better for layers of large image sizes and large channel numbers, where GEMM operations can be performed on large block sizes on matrices generated by im2col.
During iterative DNN training process, for layers can be implemented with two methods, swCaffe can run first two iterations to determine the best strategy used for remaining iterations.

\begin{figure}[ht!]
\centering
\includegraphics[width=0.45\textwidth]{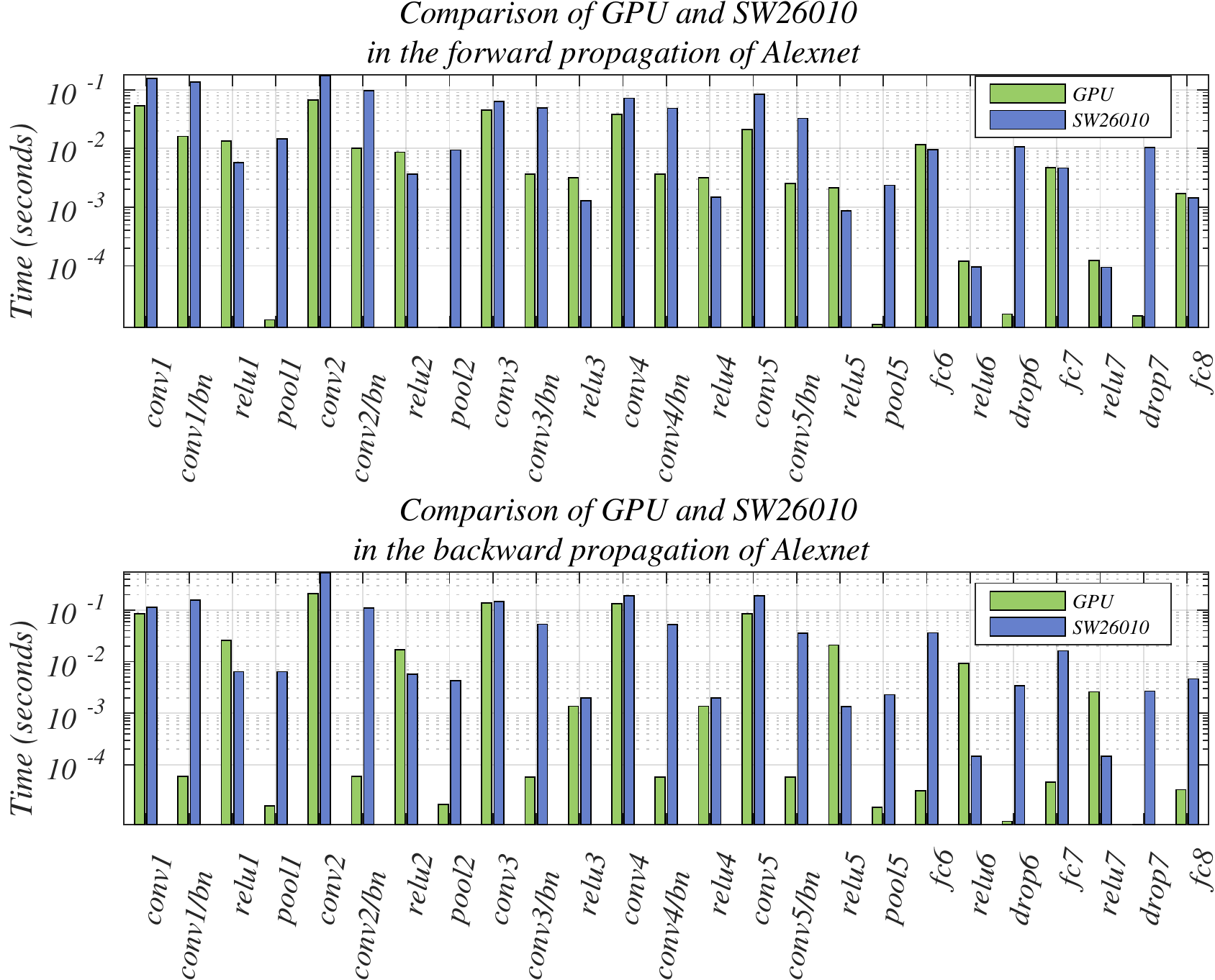}
\caption{Forward and Backward Time of Alexnet  on GPU K40m and SW26010.}
\label{fig:alex-layers}
\end{figure}

\begin{figure}[ht!]
\centering
\includegraphics[width=0.45\textwidth]{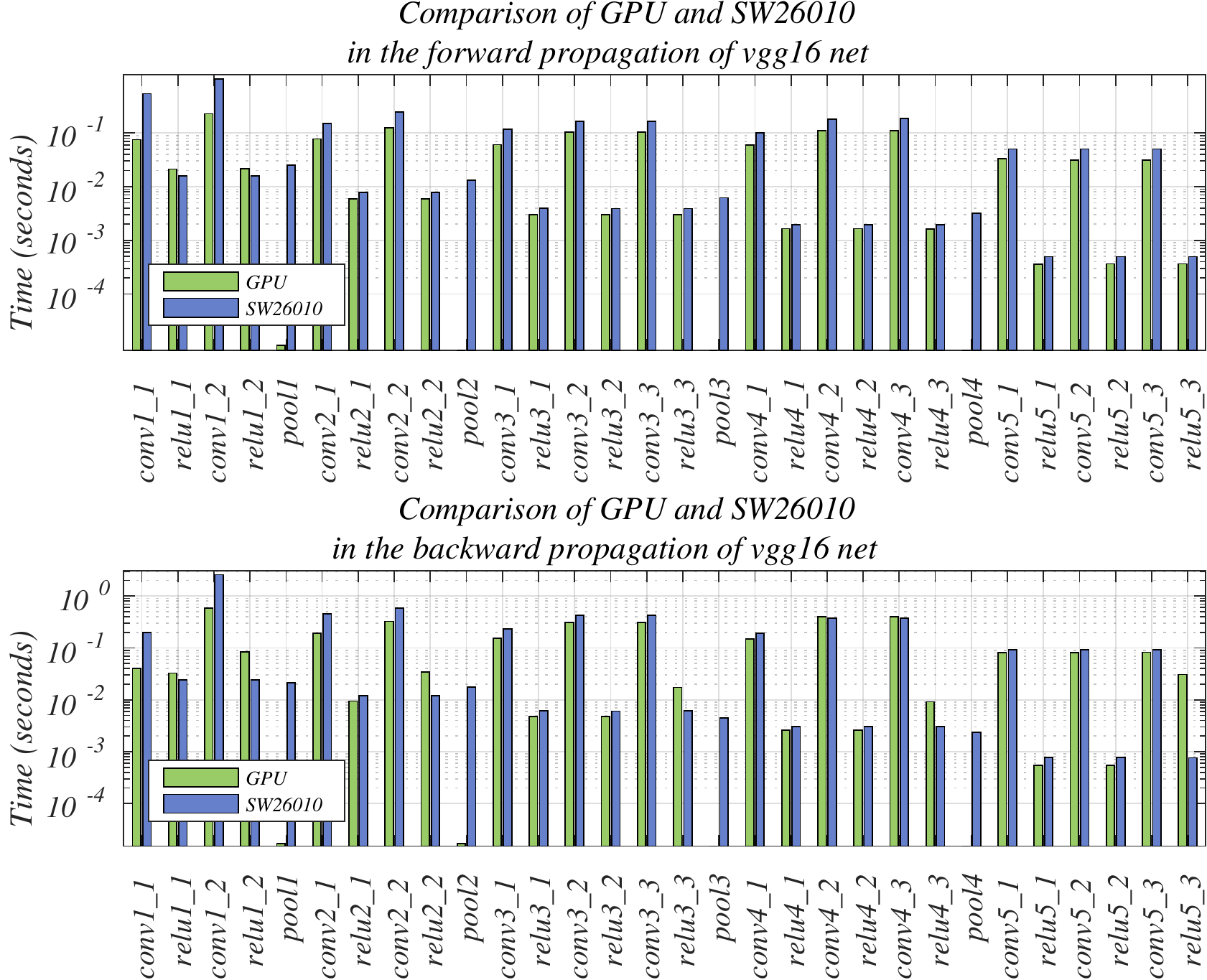}
\caption{Forward and Backward Time of VGG-16 on GPU K40m and SW26010.}
\label{fig:vgg-layers}
\end{figure}

Figure \ref{fig:alex-layers} and Figure \ref{fig:vgg-layers} present the processing time for each DNN layer on SW26010 and GPU K40m for forward propagation and backward propagation on AlexNet \cite{krizhevsky2012imagenet} and VGG-16, respectively.
We adopt some refinements to AlexNet without affecting the accuracy by changing the local response normalization (LRN) to batch normalization (BN) in AlexNet.
The performance differences between the two architectures mainly come from the following aspects
i) Although DNN training has long been considered as a compute-intensive task on GPU, we notice that most of DNN training time is spent under bandwidth-bounded situation on SW26010.
As memory bandwidth of GPU device memory can reach 288 GB/s, bandwidth-bounded layers, such as pooling layers, can be processed in device memory very fast.
However, these layers still have a significant amount of time on SW26010.
ii) Although we achieve comparative performance for most of compute-insensitive layers,
for the first two convolutional layers in both networks, SW26010 has low efficiency compared with GPU.
Given that these layers have large image sizes, im2col and col2im operations account for most of time in first two layers.
In addition, the input/output channel sizes are 3/64 and 64/64 for first two convolutional layers, which is not enough for compute-bounded blocked GEMM operations.
The flop-to-byte ratio of GEMM operation with $A$(size of $m,n$)+=$B$(size of $m,k$) $\times$ $C$(size of $k,n$) is $\frac{2mnk}{4nm + 4nk + 4mk}$.
The best ratio is $\frac{m}{6}$, if $m=n=k$.
The architectural flop-to-byte ratio calculated with the best measured bandwidth is $ratio = \frac{742.4}{28} = 26.5$.
As a result, to make GEMM be compute-bounded, we have to make $m > 160$.
However, small channel size limits the $m$ dimension sizes in transformed matrices.

\subsection{Results for different network structures}
In Table \ref{tab:perf}, we evaluate the performance of our framework on complete DNN training tasks with different network structures.
We use img/sec as an indicator, which indicates the number of images processed in one seconds.
AlexNet, VGG-16, VGG-19 \cite{simonyan2014very}, ResNet-50 \cite{he2015deep} and GoogleNet \cite{szegedy2015going} are tested with batch size as 256, 64, 64, 32, 128, respectively.
Compared with 12-core CPU, SW26010 with our framework is 3.04x\~{}7.84x faster on five DNNs.
Our framework on SW26010 outperforms K40 GPU on AlexNet with a speedup of 1.19x.
Data reading from CPU host memory to GPU device memory through PCI-E bus accounts for over 40\% time during training of AlexNet, as calculation time is too short to hide memory I/O overhead.
In contrast, CPEs in SW26010 can directly access memory with DMA so as to eliminate data reading overhead.
Our framework on SW26010 achieves 45\% and 49\% overall performance compared with NVIDIA K40m GPU on AlexNet, VGG-16, but with a theoretical memory bandwidth only 44\% of that of GPU.
Implementations of ResNet-50 and GoogleNet with swCaffe achieve 21\% and 23\% overall performance of GPU Caffe, because their convolutional layers adopt smaller channel settings than VGG-16 and VGG-19.
Since limited memory bandwidth achieved on convolutional layers with small channel numbers, the two networks exhibit stronger memory-bounded properties on SW26010.

\begin{table}[ht!]
\small
\caption{The performance (img/sec) of three processors on different DNN networks}
\begin{center}
\begin{tabular}{|c||c|c|c|c|c|c|c|}
\hline
& CPU	& NV K40m	& SW & SW/NV & SW/CPU\\
\hline
\hline
AlexNet &12.01 & 79.25 & 94.17 & 1.19 & 7.84  \\
\hline
VGG-16  & 1.06 & 13.79 & 6.21& 0.45 & 5.13\\
\hline
VGG-19   & 1.07 & 11.2 &	5.52 & 0.49 & 5.15  \\
\hline
ResNet-50  & 1.99 & 25.45 & 	5.56 & 0.21 & 2.79 \\
\hline
GoogleNet  &	4.92	& 66.09 & 14.97 & 0.23 & 3.04 \\
\hline
\end{tabular}
\end{center}
\label{tab:perf}
\end{table}

\subsection{Results for scalability}
\begin{figure}[ht!]
\centering
\includegraphics[width=0.35\textwidth]{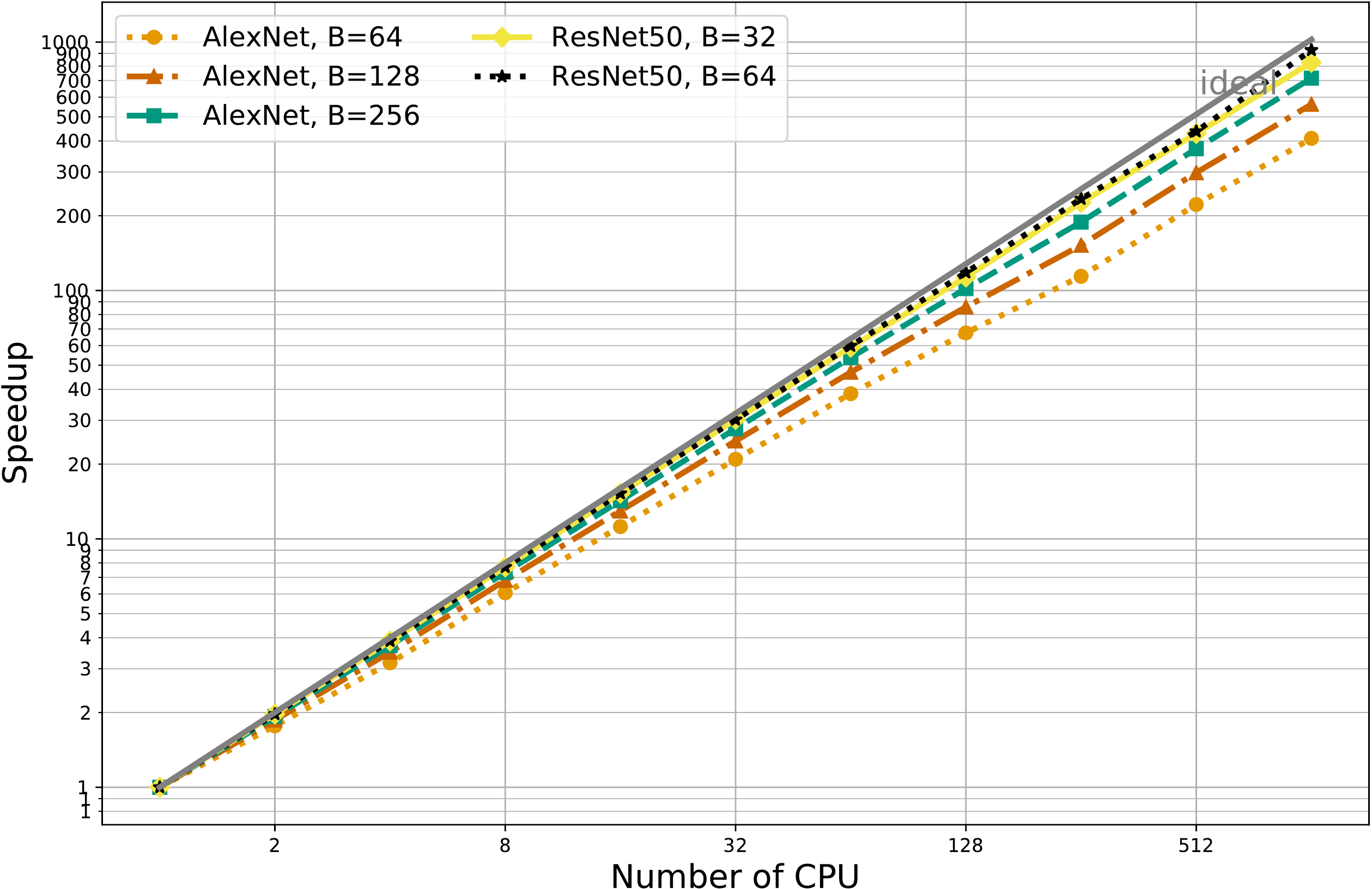}
\caption{Scalability of swCaffe.}
\label{fig:sw-scalability}
\end{figure}

Recently, works in \cite{goyal2017accurate},\cite{you2017scaling} have increased the mini-batch size in data-parallel SGD without losing accuracy over a fixed number of epochs.
Large mini-batch size can lead more possible parallelism for DNN scaling on multiple nodes, as computing task of each node can achieve high compute-to-communication ratio.
As shown Figure \ref{fig:sw-scalability}, we scale the AlexNet and ResNet50 to 1024 CPUs. 
Compared with training speed on a single node, 715.45x, 561.58x and 409.50x speedups with 1024 nodes are achieved for AlexNet trained with sub-mini-batch size as 256, 128, 64, respectively.
928.15x and 828.32x speedups with 1024 nodes are achieved for ResNet50 trained with sub-mini-batch size as 32 and 64, respectively.
Although the limit mini-batch-size of the current large-batch method \cite{you2017scaling} for AlexNet and ResNet is 32K, TaihuLight equipped our framework is able to benefit from new training algorithm with larger batch-size.

\begin{figure}[ht!]
\centering
\includegraphics[width=0.35\textwidth]{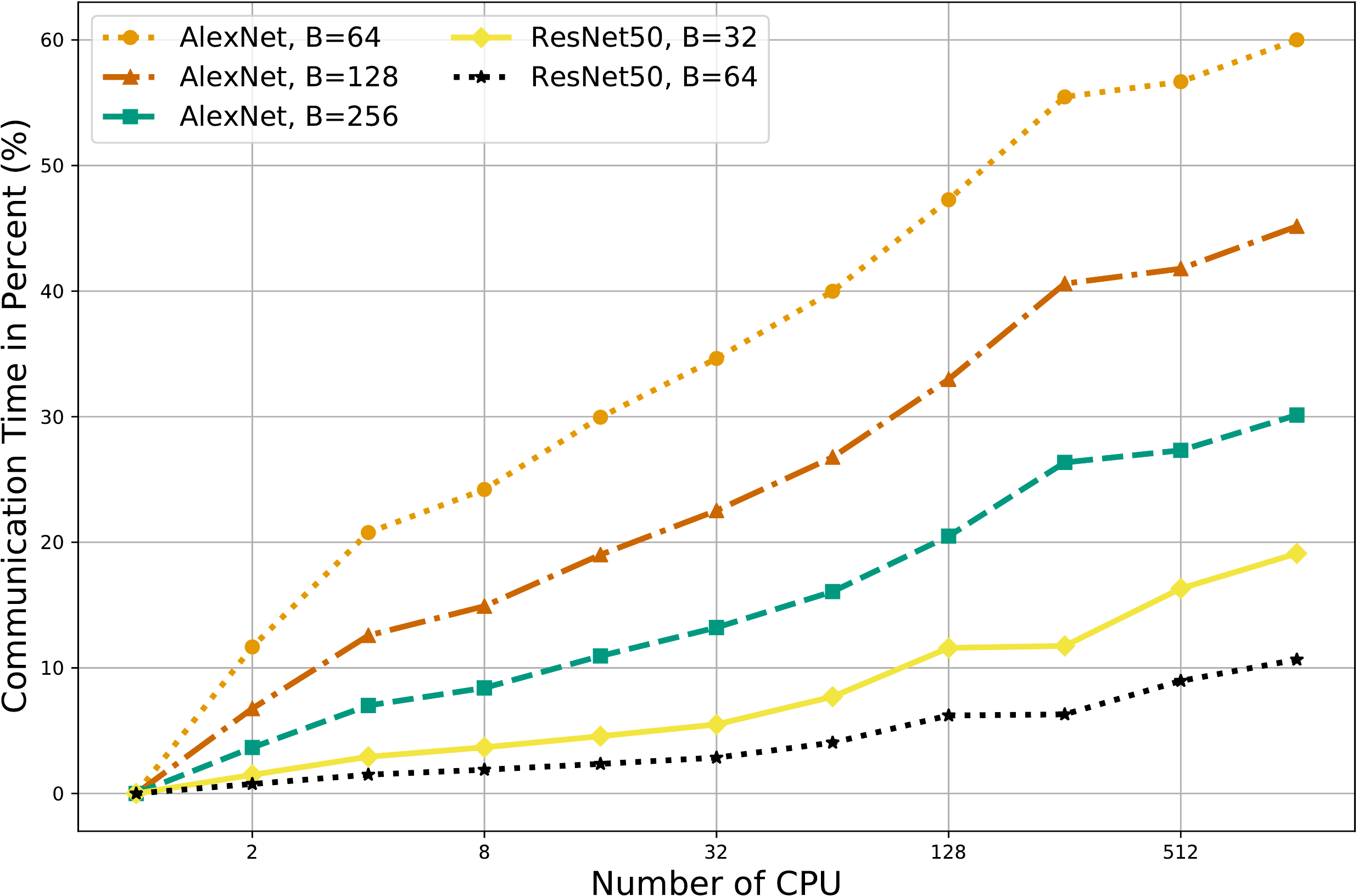}
\caption{Efficiency of swCaffe.}
\label{fig:sw-efficiency}
\end{figure}


Fig. \ref{fig:sw-efficiency}  shows the proportion of communication time during training on AlexNet and ResNet-50.
The proportion of communication time is 10.65\% and 19.11\% for ResNet-50 trained with sub-mini-batch as 32 and 64 on the scale of 1024 nodes.
The proportion of communication time is 60.01\%, 45.15\% and 30.13\% for AlexNet trained with sub-mini-batch as 64, 128, 256 on the scale of 1024 nodes.
Since the model parameter size of ResNet-50 is less than AlexNet (97.7 MB vs 232.6 MB) and more computation required for ResNet-50,
high computation-to-communication ratio accounts for better scalability of ResNet-50.

\section{Related Works}
Existing methods on accelerating basic DNN layers are mainly focused on many-core architectures of NVIDIA GPU and Intel Xeon Phi.
Library cuDNN \cite{chetlur2014cudnn} is a widely used GPU-accelerated library of primitives for deep neural networks.
Intel MKL-DNN \cite{intelcaffe} is a library of DNN performance primitives optimized for Intel architectures.
They both provide a set of highly optimized building blocks intended to accelerate compute-intensive parts of deep learning applications.

The work in \cite{catanzaro2013deep} was first proposed to train DNN models on a CPU-GPU hybrid HPC systems.
Since then, a large number of works have already been focused on scaling DNN on GPU supercomputers and HPC clusters.
Inspur-Caffe \cite{mpicaffe} is an MPI-based Caffe fork that exploits parameter-server approach with stale asynchronous gradient updates.
FireCaffe \cite{iandola2016firecaffe} discusses scaling of DNN models on a cluster of 128 GPUs connected with Infiniband interconnects.
It also adopts a allreduce-based parameter synchronization implemented with reduction trees.
S-Caffe \cite{awan2017s}  provides modern multi-GPU clusters with a CUDA-Aware MPI runtime for reducing/broadcasting operations and scales DNN training to 160 GPUs.

There are a variety of general DNN frameworks deployed on HPC systems.
Tensorflow \cite{abadi2016tensorflow} developed by Google is the most famous DNN framework that operates at large scale and in heterogeneous environments.
It implements communication using the Google RPC library.
Caffe2 \cite{caffe2} is developed by Facebook and built based on Caffe.
CNTK \cite{seide2016cntk}  developed by Microsoft.
Both Caffe2 and CNTK natively support MPI for inter-node communications.
MXNet \cite{chen2015mxnet} support multi-GPU training with a parameter server called PS-lite implemented with ZeroMQ library for communication.
Intel-Caffe \cite{intelcaffe} can harness the power of Intel KNL coprocessors and supports multi-node training by Intel MLSL (Machine Learning Scaling Library), which is a library built on top of MPI and works across various interconnects, like Intel Omni-Path, InfiniBand, and Ethernet.

\section{Conclusion}
\label{sec:conlusion}
We share our experience on designing a parallel DNN framework called swCaffe on Sunway TaihuLight from processor architecture and networking perspective.
Highly optimized routines for DNN layers are derived, fully taking into consideration different aspects of hardware characteristics.
We optimize the all-reduce operation for parameter synchronization in parallel training process in terms of both the communication topology and the computational approach.
Compared to Caffe on NVIDIA K40m GPU, our framework on SW26010 has competitive performance for DNNs with compute-intensive convolution operations, such as AlexNet and VGG.
Experimences prove our all-reduce routine is sufficient for parallel synchronous SGD training.

\bibliographystyle{unsrt}
\bibliography{sample-bibliography}
\end{document}